\documentclass[aps,prd,twocolumn,showpacs,preprintnumbers,superscriptaddress,nofootinbib]{revtex4}
\usepackage[dvips]{graphicx}
\usepackage{bm,latexsym,amsmath,amssymb,amsfonts}

\begin{document}

\title{The Riemannian Penrose inequality and a virtual gravitational collapse}

\author{Seiju Ohashi}
\affiliation{Department of Physics, Tokyo Institute of Technology, Tokyo 152-8551, Japan}
\author{Tetsuya Shiromizu}
\affiliation{Department of Physics, Kyoto University, Kyoto 606-8502, Japan}
\author{Sumio Yamada}
\affiliation{Mathematical Institute, Tohoku University, Sendai 980-8578, Japan}

\begin{abstract}
We reinterpret the proof of the Riemannian Penrose inequality by H. Bray. 
The modified argument turns out to have  
a nice feature so that the flow of Riemannian metrics appearing Bray's proof 
gives a Lorentzian metric of a spacetime. 
We also discuss a possible extension of our approach to charged black holes. 
\end{abstract}


\maketitle

\section{Introduction}

The issue of cosmic censorship is still an unsolved problem. Closely related to 
this, Penrose proposed the following inequality for the black hole \cite{Penrose}
%
\begin{eqnarray}
{\sqrt {A/16\pi}} \leq m,
\end{eqnarray}
%
where $A$ is the area of the horizon and $m$ is the ADM mass for an
asymptotically flat spacetime. 
This inequality is also yet to be proved and remains an important problem. 

In a Riemannian/time-symmetric space, Huisken and Ilmanen proved this inequality 
where the area $A$ is that of a single black hole by using the inverse mean curvature 
flow \cite{IMCF}. 
At almost the same time, Bray proved it for multi black holes using a 
conformal flow method \cite{Bray}. For the general, non-time-symmetric case, the Penrose inequality is 
still an open question. 

As we review in the next section, Bray's proof is a bit of a 
mystery. This is because it is difficult to have a physical reasoning why 
the proof works. In this paper, we introduce a normalised 
conformal flow and then we regard it as a model of the time evolution,
formulating a Lorentzian metric. 
As a result, we have a rather natural interpretation of Bray's proof. We also discuss 
some implications of our line of reasoning to the charged black hole case. 

The rest of this paper is organized as follows. In the next section, we 
review Bray's proof. In Sec. III, we present the modified proof. Then we 
give some physical interpretations of our new proof in Sec. IV. As an 
extension, we discuss the Penrose inequality for charged black holes in Sec. V. 
Finally we summarize our results in Sec. VI. 

\section{Brief sketch of Bray's proof}

We consider a time-symmetric initial data $(\Sigma, q_0)$ where $q_0$ is a 
Riemannian metric. The time-symmetric 
initial data is defined by a hypersurface in a spacetime with the zero extrinsic curvature. 
We suppose that the apparent horizons $H_0$ exist in the spacetime. It is known that the apparent 
horizon corresponds to the minimal surface in $(\Sigma, q_0)$. 

We introduce the following conformal transformation 
%
\begin{eqnarray}
q_t=u_t^4 q_0 \label{cf}
\end{eqnarray}
%
and define $v_t$ as the ``time" derivative of $u_t$ 
%
\begin{eqnarray}
v_t=\dot u_t,
\end{eqnarray}
%
where dot stands for the derivative with respect to the parameter $t$. 
We then require that $v_t$ is a harmonic function with respect to 
$q_0$ 
%
\begin{eqnarray}
\Delta_{q_0}v_t=0
\end{eqnarray}
%
with the boundary condition
%
\begin{eqnarray}
v_t(x)|_{H_t}=0 
\end{eqnarray}
%
and 
%
\begin{eqnarray}
v_t \to -e^{-t}~~{\rm as}~~r \to \infty. 
\end{eqnarray}
%

We require that $H_t$ is the minimal surface in  $(\Sigma, q_t)$. From the definition 
of $v_t$, we have 
%
\begin{eqnarray}
u_t=1+\int_0^t v_s(x)ds \to e^{-t}~~({\rm as}~r \to \infty).
\end{eqnarray}
%
Now we have a conformal flow defined by the sequence of $(\Sigma, q_t, H_t)$. 

In this conformal flow, we can show that 
%
\begin{eqnarray}
\dot A_t=0 \label{area}
\end{eqnarray}
%
and
%
\begin{eqnarray}
\dot m_t \leq 0. \label{mass}
\end{eqnarray}
%
Here $A_t$ is the area of $H_t$ and $m_t$ is the ADM mass 
for $(\Sigma, q_t)$. When we show Eq. (\ref{mass}), 
an idea of Bunting and Masood-ul-Alam 
\cite{BM} was used in a crucial way. From these we have 
%
\begin{eqnarray}
A_\infty =A_0
\end{eqnarray}
%
and
%
\begin{eqnarray}
m_\infty \leq m_0. 
\end{eqnarray}
%

In the limit of $t=\infty$, we can also show that 
$(\Sigma, q_t)$ becomes the Schwarzschild slice. Therefore 
${\sqrt {A_\infty/16\pi}}
=m_\infty$ holds. Thus, 
%
\begin{eqnarray}
{\sqrt {A_0/16\pi}} = {\sqrt {A_\infty/16\pi}}
=m_\infty \leq m_0
\end{eqnarray}
%
is proven. This is the Riemannian Penrose inequality. 

It is difficult to see why this proof works. So we will 
modify the  proof which is just a reformulation of the 
conformal flow. Although the new argument requires rather minor 
technical modifications from Bray's one, we gain a new insight,
which in turn offers a physical interpretation to the conformal flow. 

\section{Normalized conformal flow}

Let us introduce the following conformal transformation 
%
\begin{eqnarray}
\tilde{q}_t =\tilde{u}^4_t q_0,
\end{eqnarray}
%
where $\tilde u_t$ is defined by 
%
\begin{eqnarray}
\tilde u_t = \Bigl(\frac{m_0}{m_t}\Bigr)^{1/2}u_t. 
\end{eqnarray}
%
$u_t$ is the same with the previous one in Eq. (\ref{cf}). 
Now we have a new flow $(\Sigma, \tilde q_t, H_t)$. Note that
the surface $H_t$ remains minimal after the dilation of the metric. 
It is easy to show 
%
\begin{eqnarray}
\dot{\tilde{m}}_t=0.
\end{eqnarray}
%
In addition, 
%
\begin{eqnarray}
\dot{\tilde{A}}_t =4 \int_{H_t} (\dot{{\tilde u}}_t / \tilde{u}_t) dS.
\end{eqnarray}
%
In the integrand, 
%
\begin{eqnarray}
\dot{\tilde{u}}_t=\Bigl(\frac{m_0}{m_t}\Bigr)^{1/2}v_t
-\frac{1}{2}\Bigl(\frac{m_0}{m_t}\Bigr)^{1/2}\frac{\dot m_t}{m_t}u_t.
\end{eqnarray}
%
Since $\dot m_t \leq 0$
%
\begin{eqnarray}
{\dot {\tilde u}}_t|_{H_t}=-\frac{1}{2}\Bigl(\frac{m_0}{m_t}\Bigr)^{1/2}
\frac{\dot m_t}{m_t}u_t|_{H_t} \geq 0.
\end{eqnarray}
%
Thus
%
\begin{eqnarray}
\dot{\tilde{A}}_t \geq 0. 
\end{eqnarray}
%
We can show that the space becomes the Schwarzschild slice in the $t=\infty$ limit 
as well as the case of the conformal flow. Thus, 
$16\pi \tilde{m}_\infty^2=\tilde{A}_\infty$ holds. Finally we can show 
the Riemannian Penrose inequality again as 
%
\begin{eqnarray}
16\pi m_0^2=16\pi \tilde m_\infty^2=\tilde A_\infty \geq A_0. 
\end{eqnarray}
%

Namely over this normalized conformal flow, the ADM mass is conserved and 
the area of the apparent horizon is increasing. The former corresponds to 
the well-known fact that the ADM mass is a conserved quantity in asymptotically 
flat spacetimes. 
The latter corresponds 
to the area theorem of black holes (See Ref. \cite{Hayward} for the area theorem of 
apparent horizon). These features offers 
a nice physical interpretation of the normalized conformal flow. 
In the next section, we will look at this context more closely.

\section{Physical Interpretation}

\subsection{General spacetime}

From now on, we will regard the normalised conformal flow as a time evolution. 
We suppose that the time evolution is given by 
%
\begin{eqnarray}
ds^2 & = & g_{\mu\nu}dx^\mu dx^\nu
=-\alpha^2(t,x)dt^2+\tilde{q}_t \nonumber \\
& = & -\alpha^2(t,x)dt^2+\tilde{q_t}_{ij}dx^i dx^j,
\end{eqnarray}
%
where $\alpha$ is the lapse function and $\tilde{q_t}_{ij}$ is the component of $\tilde{q_t}$. 
Later we will choose $\alpha$ so that the $t=$ const. slices are asymptotically 
flat in the usual sense. In this case the extrinsic curvature of $t=$const. hypersurfaces becomes 
%
\begin{eqnarray}
K_{ij}=\frac{1}{2\alpha}\partial_t {\tilde{q_t}}_{ij}=
2\frac{\dot{\tilde{u}}_t}{\alpha \tilde{u}_t}  {\tilde{q_t}}_{ij}.
\end{eqnarray}
%
Then it turns out that the expansion rate $\theta$ of the outgoing null geodesic 
congruence on $H_t$, (which is by definition, equal to $h^{\mu\nu} \nabla_\mu (t_\nu+r_\nu)$
where $r^\mu$ is the unit normal vector to $H_t$ in $(\Sigma, \tilde{q}_t)$, $t^\mu$ is the 
unit coordinate vector, making $t^\mu+r^\mu$ outgoing null vector, $h_{\mu\nu}$ is the induced metric
on the surface $H_t$,) is non-negative
%
\begin{eqnarray}
\theta|_{H_t} \propto (k+ K- K_{ij}r^ir^j)|_{H_t}
=-2\frac{\dot{m}_t}{\alpha m_t} \geq 0.
\end{eqnarray}
%
This is because of $\dot{m}_t \leq 0$(See Eq. (\ref{mass})). Here $k$ is
the trace of extrinsic curvature of $H_t$ with respect to $\tilde{q}_t$ and 
$K=K^i_{i}$. 
Thus $H_t$ is located outside an apparent horizon/marginally trapped surface 
in a virtual spacetime $(M,g)$. 

In the time evolution of $H_t$, we can see that $H_t$ approaches to the 
apparent horizon 
%
\begin{eqnarray}
\theta|_{H_t} \propto -2\dot m_t/m_t \to 0,
\end{eqnarray}
%
because we know that the final state at $t=\infty$ is Schwarzschild slice, 
the convergence implies $\dot{m}_t \to 0$ as $t \to \infty$. 

Let us suppose that $(M,g)$ satisfies the four dimensional Einstein 
equation
%
\begin{eqnarray}
{R}_{\mu\nu}-\frac{1}{2}{g}_{\mu\nu}R=8\pi T_{\mu\nu},
\end{eqnarray}
%
where ${R}_{\mu\nu}$ and $R$ are the Ricci curvature and scalar curvature of
 $g$. Here we do not yet have the above equation determining  the virtual spacetime.  
The stress tensor $T_{\mu\nu}$ needs to be chosen so that the above equation is satisfied.  

To do so, let us focus on the Hamiltonian and momentum constraints, 
%
\begin{eqnarray}
{}^t \tilde{R}+ K^2- {K}_{ij}  K^{ij}=16\pi \rho \label{Hami}
\end{eqnarray}
%
and
%
\begin{eqnarray} \label{Moment}
\tilde D^i  {K}_{ij}-\tilde D_j K
=-8\pi J_j,
\end{eqnarray}
%
where $\rho=T_{\mu\nu}t^\mu t^\nu$, $J_i=T_{\mu i}t^\mu$. ${}^t \tilde{R}$ and $\tilde{D}_i$ are 
the Ricci scalar the covariant derivative with respect to $\tilde{q}_t$, 
respectively. From the 
Hamiltonian constraint, we can calculate $\rho$ 
%
\begin{eqnarray}
16\pi \rho = 16\pi \Bigl( \frac{m_t}{m_0}\Bigr)^2u_t^{-4}\rho_0
+24\frac{1}{\alpha^2} \Bigl(  \frac{\dot{\tilde{u}}_t}{\tilde{u}_t} \Bigr)^2 \geq 0. 
\label{density}
\end{eqnarray}
%
In the above we used ${}^0 \tilde R=16\pi \rho_0$, where $\rho_0$ is the energy 
density of real matters in the physical initial data. Note that $\rho_0$ is not one computed 
from virtual matters $T_{\mu\nu}$ here. 
Then we see that $\rho$ comes out to be non-negative. This is a nice feature in 
the physical sense. 

Next we can calculate $J_i$ and the result is 
%
\begin{eqnarray}
2\pi J_i=\partial_i(v_t/\alpha u_t).
\end{eqnarray}
%
On $H_t$, we have 
%
\begin{eqnarray}
2\pi J_i|_{H_t}=\partial_iv_t/(\alpha u_t)|_{H_t}.
\end{eqnarray}
%
Since $v_t(x)$ is the harmonic function, the maximum principle 
tells us $\partial_iv_t \leq 0$ outward direction of $H_t$. More 
precisely, if one introduces the outward normal vector $r^i$ of $H_t$ in 
$t=$const. slices, $r^i \partial_iv_t \leq 0$.  Thus 
we can see the ingoing energy flux of artificial matters, that is, 
$r^i J_i \leq 0$.

Here note that $K=6\dot{\tilde{u}}_t/\alpha \tilde{u}_t \simeq 6v_t/\alpha u_t 
\to -6/\alpha$ as $r \to \infty$. If $\alpha$ is taken to be $\sim r^2$ at $r=\infty$, 
we can make $t=$const. slice to be asymptotically flat in the usual way. 
In the energy density of the virtual matter, the second term of right-hand side 
of Eq. (\ref{density}) is proportional to $r^{-4}$. So it behaves like a radiation. 
Here we note that $T_{ij}$ is determined by the following algebraic equation
for $T_{ij}$.  Note that everything else has been already picked.
%
\begin{eqnarray}\label{T-tensor}
& & -\tilde D_i \tilde D_j \alpha+
\alpha({}^t \tilde{R}_{ij}+{K}^k_k {K}_{ij}-2{K}_{ik}{K}^k_j)
+\dot {K}_{ij} \nonumber \\
& & =
8\pi \alpha \Bigl(T_{ij}+\frac{1}{2} g_{ij}(\rho-T^k_k) \Bigl),
\end{eqnarray}
%
where ${}^t \tilde {R}_{ij}$ is the Ricci tensor with respect to 
$\tilde{q}_t$. Using this, in principle, we can check if the dominant 
energy condition is satisfied. However, we have to compute the 
second derivative of $u_t$, which is included in $\dot {K}_{ij}$, 
to do so. Unfortunately, the information of the second derivative is 
not given in the normalised conformal flow. Thus it is difficult to 
see if the dominant energy condition is satisfied. We would expect 
that we can choose the lapse function, $\alpha$, so that the 
dominant energy condition is satisfied. This issue is beyond of 
current work. 

As a consequence, we have the following physical picture for the 
normalised conformal flow. The virtual time evolution corresponds to 
the gravitational collapse. From the behavior of virtual matters characterized
by $T_{\mu\nu}$, the 
3-dimensional hypersurface $\cup_t H_t$ looks like a horizon. Moreover, the area of 
$H_t$ is increasing with time. We recall in Bray's construction that 
the topological type of the surface $H_t$ may change, as the 
surface may jump across some singular times.
And $H_t$ approaches to the horizon because the expansion 
rate of null congruence on $H_t$ 
is decaying to zero at $t=\infty$. Thus, the normalized conformal flow 
gives us a virtual gravitational collapse. Since the final state 
is promised to be Schwarzschild slice in this evolution, it is natural 
to have the Penrose inequality. If we know that the final state is 
Schwarzschild slice, the area theorem implies the Penrose inequality. 

\subsection{Example: evolving Schwarzschild slice}

As an example, we consider the virtual spacetime modeled by the normalised 
flow for the Schwarzschild slice.
%
\begin{eqnarray}
ds^2 & = &-\alpha^2dt^2+\tilde {q}_{tij}dx^idx^j \notag \\
& = & -\alpha^2dt^2+\left( e^{-t}+\frac{M}{2r}e^t\right) ^4\left( dr^2+r^2d\Omega ^2\right) .
\end{eqnarray}
%
Here $H_t$ is located at $r=\frac{M}{2}e^{2t}$. Note that the mass $2(e^{-t})(M e^t/2)$ and the area of $H_t=16\pi M^2$ are both kept 
constant in $t$.  Furthermore,  
we would emphasize that the above spacetime is not obtained by  
a coordinate change of the Schwarzschild {\it spacetime} metric and 
in particular does not satisfy the vacuum Einstein equation.
Instead, it will be made to satisfy the non-vacuum Einstein equation 
{\it driven by} a suitably chosen Eq. (\ref{T-tensor}) stress-energy tensor $T_{\mu\nu}$ as seen below.

The extrinsic curvature of $t=\text{const}.$ hypersurface is
%
\begin{equation}
K_{ij}=\frac{2}{\alpha } \Biggl( \frac{-e^{-t}+\frac{M}{2r}e^t }{ e^{-t}+\frac{M}{2r}e^t }
\Biggr) \times \tilde {q}_{tij}.
\end{equation}
%
On $H_t$, we see $K_{ij}|_{H_t}=0$. This is a peculiar feature for 
the normalised conformal flow of the Schwarzschild spacetime. Note that our
normalization is trivial for $\dot{m}=0$ in this evolution. This indicates that the 
horizon does not have a nontrivial time evolution in the current virtual dynamical evolution. 
Indeed, we can check that that 
the expansion rate of outgoing null geodesic congruence $\theta_t$ vanishes as 
%
\begin{equation}
\theta |_{H_t}\propto \left( k+K-K_{ij}r^ir^j \right) |_{H_t}=0.
\end{equation}
%
Here we used the fact that $H_t$ is the minimal surface and  $K_{ij}|_{H_t}=0$. 
This means $H_t$ coincide with the apparent horizon of the virtual gravitational 
collapse throughout the evolution . 

From the Hamiltonian constraint of Eq. (\ref{Hami}), we can see that the matter density on $H_t$ vanishes as 
%
\begin{equation}
16\pi \rho |_{H_t}=\left( {}^t \tilde {R}+K^2-K_{ij}K^{ij}\right) |_{H_t}=0.
\end{equation} 
%
In the above we used ${}^t \tilde {R}=0$. On the other hand, the 3-momentum of Eq. (\ref{Moment}) is evaluated as 
%
\begin{eqnarray}
J_r|_{H_t}& = & -\frac{1}{8\pi }\left( D^jK_{jr}-D_rK\right) |_{H_t} \notag \\ 
   & = & -\frac{e^{-2t}}{2\pi \alpha M} \leq 0.
\end{eqnarray}
%
Thus we see that the artificial matter represented by $T_{\mu\nu}$ has the trivial energy density
on $H_t$.  This is merely consistent with the fact that the area of $H_t$ does not increase 
with the time.  On the other hand, it has a nontrivial ingoing (through $H_t$) 3-momentum $J_r$.

\section{Implication to charged black holes}

Although our new proof is just a rearrangement of Bray's proof, 
there is a possibility to apply it to other issues. For example, 
one may want to address the Penrose inequality for charged 
black holes. According to Ref. \cite{YW}, Bray's argument is hoped to
be generalized so that 
%
\begin{eqnarray}
m_0 \geq m_\infty= \frac{1}{2}
\Bigl(R+\frac{Q^2}{R} \Bigr)
\end{eqnarray}
%
holds where $Q$ is the charge of black holes. The Reissner-Nordstr\"{o}m slice 
realizes the equality. Introducing the area 
radius by $R={\sqrt {A_0/4\pi}}={\sqrt {A_\infty/4\pi}}$, the above 
is rewritten by 
%
\begin{eqnarray}
m_0-{\sqrt {m_0^2-Q_0^2}} \leq R \leq m_0+{\sqrt {m_0^2-Q_0^2}}. 
\end{eqnarray}
%
However, in Ref. \cite{YW}, a counterexample to the lower 
bound was constructed. Because of the evidence, it is unlikely that Bray's proof 
works for charged black holes in the way presented above. 

On the other hand, we may expect that the 
upper bound for the area radius holds. Namely  we hope
to show that the inequality 
%
\begin{eqnarray}
4\pi \Bigl(m_0+{\sqrt {m_0^2-Q_0^2}}\Bigr)^2=A_\infty \geq A_0=4\pi R^2
\end{eqnarray}
%
(that is 
$
m_0+{\sqrt {m_0^2-Q_0^2}} \geq R )
$
holds. The lesson to be learned from the counterexample is that in Bray's
original flow, the area radius was fixed while the mass was decreased via the
flow, though physically the area should be increased till it reaches the maximal
value set by the fixed mass.  This is what we have done with the normalization.
So with charge in play, we may hope to prove
with $m$ and $Q$ fixed, the area can be increased till it reaches that of 
Reissner-Nordstr\"{o}m's specified by the parameters $(m_0,Q_0)$.

\section{Summary}

In this article, we proposed a proof of the Riemannian Penrose 
inequality which is a modification of Bray's proof (Ref.\cite{Bray}.) In the original  
proof by Bray,  a conformal flow of the Riemannian metrics was employed,
so that the mass is decreasing while
the area of the horizon is fixed. However, it is difficult to see the physical 
reason why the proof works. Hence we proposed a dual viewpoint by 
normalizing the conformal flow. It is a family of  conformal transformations
so that now the mass is fixed while the area is increasing. Then we 
observed that the behaviors of the dual flow enjoy some plausible physical 
features, that is, the normalised conformal flow corresponds to 
a virtual time evolution of gravitational collapse, satisfying a non-vacuum 
Einstein equation. In addition, our new approach may shed some new 
light to prove the following Penrose type inequality for charged black holes. 
%
\begin{eqnarray}
4\pi \Bigl(m_0+{\sqrt {m_0^2-Q_0^2}}\Bigr)^2 \geq A_0,
\end{eqnarray}
%
which is consistent with a picture (Ref.\cite{PenroseYau}) resulting from the cosmic censorship  as well as the so-called no-hair theorem where an evolving black hole is expected to settle
down to a Kerr(-Newman) spacetime with the parameters $(m_0, Q_0)$ specified by
the initial slice. This is left for future study. 


\acknowledgments

We are grateful to S. Kinoshita, R. Mizuno, N. Tanahashi and K. Tanabe for useful comments. 
TS is partially supported by Grant-Aid for Scientific Research from Ministry of 
Education, Science, Sports and Culture of Japan (Nos.~20540258 
and 19GS0219), the Japan-U.K. Research Cooperative Programs. 
SY is partially supported by Grant-Aid for Scientific Research (No.~20540201).




\end{document}